\documentclass{SolarPhysics}    % Specifies the document style.
\usepackage[optionalrh]{spr-sola-addons}
\usepackage{epsfig}
\usepackage{url}
\usepackage{fixltx2e}
\usepackage[usenames]{color}

\newcommand{\gapprox}{\lower.4ex\hbox{$\;\buildrel >\over{\scriptstyle\sim}\;$}}
\newcommand{\lapprox}{\lower.4ex\hbox{$\;\buildrel <\over{\scriptstyle\sim}\;$}}
\newcommand{\arcsec}{\hbox{$^{\prime\prime}$}}

\def\etal{{\it et al.,~}}
\def\eg{{\it e.g.}\ }
\def\ie{{\it i.e.}\ }

\begin{document}
\begin{article}
\begin{opening}
\title{ 	Blind Stereoscopy of the Coronal Magnetic Field 	}

\author{        Markus J. Aschwanden$^1$ $\bullet$ Carolus J. Schrijver$^1$  
		$\bullet$ Anna Malanushenko$^2$}

\runningauthor{M.J. Aschwanden \textit{\etal}}
\runningtitle{Blind Solar Stereoscopy}

\institute{$^1)$Lockheed Martin, 
		Solar and Astrophysics Laboratory, 
                Org. A021S, Bldg.~252, 3251 Hanover St.,
                Palo Alto, CA 94304, USA;
                e-mail: aschwanden@lmsal.com, schrijver@lmsal.com }

\institute{$^2)$ University Corporation for Atmospheric Research (UCAR),
		P.O.Box 3000, 
		Boulder, CO 80307, USA} 

\date{Received 7 May 2015; Revised 4 June 2015; Accepted 15 June 2015}

\begin{abstract}
  We test the feasibility of 3D coronal-loop tracing in stereoscopic
  EUV image pairs, with the ultimate goal of enabling efficient 3D
  reconstruction of the coronal magnetic field that drives flares and
  coronal mass ejections (CMEs). We developed an automated code
  designed to perform triangulation of coronal loops in pairs (or
  triplets) of EUV images recorded from different perspectives.  The
  automated (or blind) stereoscopy code includes three major tasks:
  (i) automated pattern recognition of coronal loops in EUV images,
  (ii) automated pairing of corresponding loop patterns from two
  different aspect angles, and (iii) stereoscopic triangulation of 3D
  loop coordinates.  We perform tests with simulated stereoscopic EUV
  images and quantify the accuracy of all three procedures. In
  addition we test the performance of the blind stereoscopy code as a
  function of the spacecraft-separation angle and as a function of the
  spatial resolution. We also test the sensitivity to magnetic
  non-potentiality. The automated code developed here can be used for
  analysis of existing {\sl Solar TErrestrial RElationship Observatory
  (STEREO)} data, but primarily serves for a design
  study of a future mission with dedicated diagnostics of
  non-potential magnetic fields. For a pixel size of 0.6\arcsec
  (corresponding to the {\sl Solar Dynamics Observatory (SDO)
  Atmospheric Imaging Assembly (AIA)} spatial resolution of 1.4\arcsec),
  we find an optimum spacecraft-separation angle of $\alpha_s
  \approx 5^\circ$.
\end{abstract}

\keywords{Sun: UV radiation --- Magnetic fields --- Methods --- Stereoscopy }

\end{opening}

\section{	Introduction				}

One critically needed tool for forecasting severe geomagnetic storms
well ahead of time is a reliable method to map the magnetic field
erupting into the heliosphere during coronal mass ejections (CMEs) so that
its evolution can be modeled well before their field impacts Earth's
(\eg Schrijver \etal 2015). A major impediment at present
is that we cannot reliably describe the magnetic field in nascent CMEs
as their erupting structure enters the heliosphere.

Even with recent advances in the ready availability of vector-magnetic
data on active regions, the modeling of the configuration of the
coronal field above these regions remains a challenge. Model results
based on surface observations alone are generally ambiguous and not
well representative of the observed coronal configuration (\eg
DeRosa \etal 2009, and references therein). The use of coronal-loop
trajectories to guide field models towards a solution compatible with
surface-field measurements shows promise (Malanushenko \etal 2012,
2014), but the fact that only the 2D trajectories projected against
the plane of the sky are available presents a stumbling block that
needs to be overcome. Malanushenko \etal (2012) approximate the
third coordinate, along the line-of-sight (LOS), by pairing up each observed
coronal loop with the best-fitting field line of a linear force-free
field (using separate field lines for each loop), and they then 
iterate towards a non-linear force-free field while continually
nudging the overall solution back to the set of 3D loop trajectories
first determined. Whereas this results in a model field that by
design matches the observed loops quite well, it is unlikely that the
input 3D trajectories are entirely correct (\eg see differences between
2D and 3D reconstructions in Aschwanden 2013b). Until we have a way to
measure the 3D loop trajectories we cannot truly validate the
method, but once the 3D trajectories are known, they can of course be
used from the outset to guide the model field towards a solution
compatible with the observed 3D configuration.

Here, we study a concept with two or three spacecraft that provide stereoscopic
views of EUV images of coronal loops, which when combined with
photospheric line-of-sight magnetograms, provide information suitable
for 3D reconstruction of the coronal magnetic field.  One possible
orbital spacecraft configuration is formed by the Sun--Earth Lagrangian
points L1 and L4 (or L5), similar as it was obtained when the 
{\sl Solar TErrestrial RElationship Observatory (STEREO) A(head)}
and {\sl B(ehind)} spacecraft moved near the L4 and L5 points in 2008, 
while the {\sl Solar and Heliospheric Observatory (SOHO)}
was positioned in L1. However, that was a very temporary
configuration, with instrumentation at moderate resolution. Ideally, a
new generation of instruments should have a spatial resolution that is
comparable to that of {\sl Solar Dynamics Observagory (SDO)}'s 
{\sl Atmospheric Imaging Assembly (AIA)}, and a field of view at least 
as large as supported by its {\it $4k\times4k$} imaging cameras with 
0.6\arcsec
pixels. Unlike the STEREO--SOHO configuration, a scientifically more
promising mission should provide a long-lived multi-spacecraft
configuration, providing images with essentially identical passband
and telescope characteristics.

We envision that a pair (or perhaps a triplet) of spacecraft equipped
with the necessary instruments could support the autonomous calculations
of the coronal magnetic field, its non-potentiality, and the free
energies in each active region either in near-real time or with up to
at most a day delay, so that this compound observatory can be used to
understand active-region instabilities and heliospheric model input, and
as an early-warning system for severe space-weather storms. To
succeed, the data-processing and modeling capabilities would require
(i) automated pattern recognition of coronal loops in EUV images, (ii)
automated stereoscopic pairing of coronal loops, (iii) stereoscopic
triangulation of coronal loops, and (iv) nonlinear force-free field
forward-fitting that yields the non-potential magnetic field, its free
energy, and (v) -~after eruptions~- quantitative information on the 
ejecta into the heliosphere. We demonstrate the feasibility of
the first four of these automated tasks in this study, and we
constrain the optimum configuration for the angular spacecraft
separation.

Recent reviews on solar stereoscopy and tomography have been presented by 
Aschwanden (2011), and recent reviews on the coronal magnetic field
are given by Wiegelmann and Sakurai (2012) and Wiegelmann, Thalmann, and
Solanki (2014). Early attempts of solar stereoscopy using information
from a single spacecraft (using XUV images from {\it Skylab)} used the solar
rotation to measure stereoscopic parallaxes (Berton and Sakurai 1985),
which requires (unrealistic) static coronal loops on time scales of at
least one day, but hydrodynamic heating and cooling processes of loops
occur on time scales of $\approx 10^3$ seconds in active regions (\eg
Warren and Winebarger 2007), even as the field's photospheric boundary
is evolving underneath.  A dynamic solar-rotation stereoscopy method
was developed later by Aschwanden \etal (1999, 2000), which relieves
the requirement of static loops {\it in lieu} of a quasi-static
magnetic field.  This assumption is somewhat more realistic, but
breaks down after about one day, since photospheric magnetic fields
involved in major flaring and eruptions were observed to have
characteristic growth and decay timescales of approximately one to two days
(Pevtsov, Canfield, and Metcalf 1994; Schrijver \etal 2005; 
Welsch, Christe, and McTiernan 2011).
Therefore, the only solution for accurate stereoscopy requires
simultaneous measurements with multiple spacecraft.  

The first stereoscopic reconstruction of coronal loops using two
simultaneous spacecraft observations was conducted with STEREO/A and B
(Feng \etal 2007; Aschwanden \etal 2008), but the tracing of coronal
loops was carried out manually, which is subject to human judgement
and does not enable efficient processing in real-time, nor in rapid
time intervals nor with large statistics.  A fully automated 
pattern-recognition algorithm that extracts the 2D geometry of coronal 
loops and performs magnetic modeling has been employed in a recent study
(Aschwanden, Xu, and Jing 2014), applied to 172 flare events in
numerous active regions. This algorithm also performed nonlinear
force-free field modeling and determined the evolution of the free
energy during flare events, using high-resolution images of SDO/AIA,
but this algorithm uses information on the 2D geometry from a single
spacecraft only, and thus is expected to retrieve less accurate
information on the magnetic field than would be possible from
stereoscopically determined 3D geometries of coronal loops
(Aschwanden 2013b).
Comparisons of NLFFF reconstructions using single-spacecraft 2D
{\it versus} dual-spacecraft 3D geometries yielded consistent results for a
simple forward-fitted quasi-NLFFF model in terms of vertical currents
(Aschwanden 2013), but the accuracy for more general NLFFF solutions
is not known. On the other hand, accurate NLFFF solutions fitted to
coronal loop geometries have been accomplished with a
Quasi-Grad-Rubin method (Malanushenko \etal 2012, 2014), but the
fitting constraints were based on manual tracing of loops and the
computation time with the present code prevents efficient real-time
calculations. Even observations from STEREO/A and B during the most
optimum conditions at small spacecraft-separation angles (during 2007)
were not able to provide accurate magnetic-loop geometries, because
the spatial resolution of the STEREO {\sl Extreme UltraViolet Imager
(EUVI)} images is too poor, being
three times poorer than SDO/AIA images.  Given all of these instrumental
and computational restrictions, an ideal multi-spacecraft
configuration suitable for most accurate magnetic-field modeling and
optimum signal-to-noise ratio has still to be established with a
new design for a future mission.

In this article we test the principle of dual and triple stereoscopy to
establish the 3D coronal loop configuration from simultaneous EUV
image combinations. In the process, we develop a suite of numerical
codes that is capable of performing stereoscopy in an automated way,
which includes simulations of synthetic stereoscopic image pairs
(section 2), automated detection of coronal loops in high-resolution
EUV images (section 3), automated stereoscopic pairing of loops
(section 4), and automated 3D triangulation of loops (section 5).  We
investigate the accuracy of stereoscopy as a function of the number of
spacecraft (section 6), as a function of the spacecraft-separation
angle (section 7), as a function of the spatial resolution (section
8), spacecraft position (section 9), and its sensitivity to the
non-potentiality of the magnetic field (section 9). Discussions and
Conclusions are provided in section 10.

\section{	Simulation of Stereoscopic Images              }

For our simulations of EUV images suitable for testing the principle of multi-spacecraft 
stereoscopy we choose data from active region NOAA 11158, as observed 
on 15 February 2011, around the time of a GOES X2.2-class flare 
event that occurred at 01:56 UT. This active region produced the
first X-class flare event in the era of the  
SDO (Pesnell, Thompson, and Chamberlin 2011), and this is one of the
best studied regions. This active region was also chosen in 
previous magnetic modeling with nonlinear force-free field (NLFFF) 
methods (Malanushenko \etal 2014), using {\sl Helioseismic and 
Magnetic Imager} (HMI) (Scherrer \etal 2012) and AIA (Lemen \etal 2012) data.

In our simulation of image sets from different perspectives, we start with 
a line-of-sight
magnetogram of HMI only, acquired at 15 February 2011, 01:40 UT. Observed from 
Earth perspective, the center of AR 11158 has a heliographic position of S21W12,
which is centered at $[x_0,y_0] \approx [0.10,-0.35] {\rm R}_\odot$
from disk center.
The HMI image has a pixel size of 0.50422$\arcsec$ and the solar radius
is 1927.2 pixels.  We extract a subimage with a field-of-view 
$x=[0.0, 0.5] {\rm R}_\odot$ and $y=[-0.5, 0.0] {\rm R}_\odot$, which corresponds 
to a size of $965 \times 965$ HMI pixels.

We decompose the magnetogram into $n_{\rm mag}=100$ sources and calculate
the potential field that results from the combined field of the 100
subphotospheric unipolar magnetic charges, according to the method
described in Aschwanden and Sandman (2010) and in Appendix A of 
Aschwanden \etal (2012b). 
The line-of-sight magnetic field component has a
range of $-700$ G $< B_z(x,y) <$ 1017 G within the chosen
field-of-view.  We cover the $965 \times 965$ pixel subimage with a
grid of $100 \times 100$, and define each grid point that has a LOS
field $B_z(x,y) \ge 200$ G as a footpoint of a coronal loop, from
which we extrapolate the potential-field line until it hits one of the
six boundary sides of the computation box, using a height of
$h_{\rm max}=0.15$ ${\rm R}_\odot$ (or 104 Mm). From the 10,000 grid points, the
magnetic field exceeds the minimum limit of 200 G at 261 locations,
which yields 261 potential-field lines. We rotate the computation box
to different viewing angles, using the same coordinate transformation
as solar rotation produces, for instance rotating by $+15^\circ$
to the West, in order to mimic a viewing position of STEREO-B at
position E15 eastward on the Earth (Figure 1, top panels).

\begin{figure}
\centerline{\includegraphics[width=1.0\textwidth]{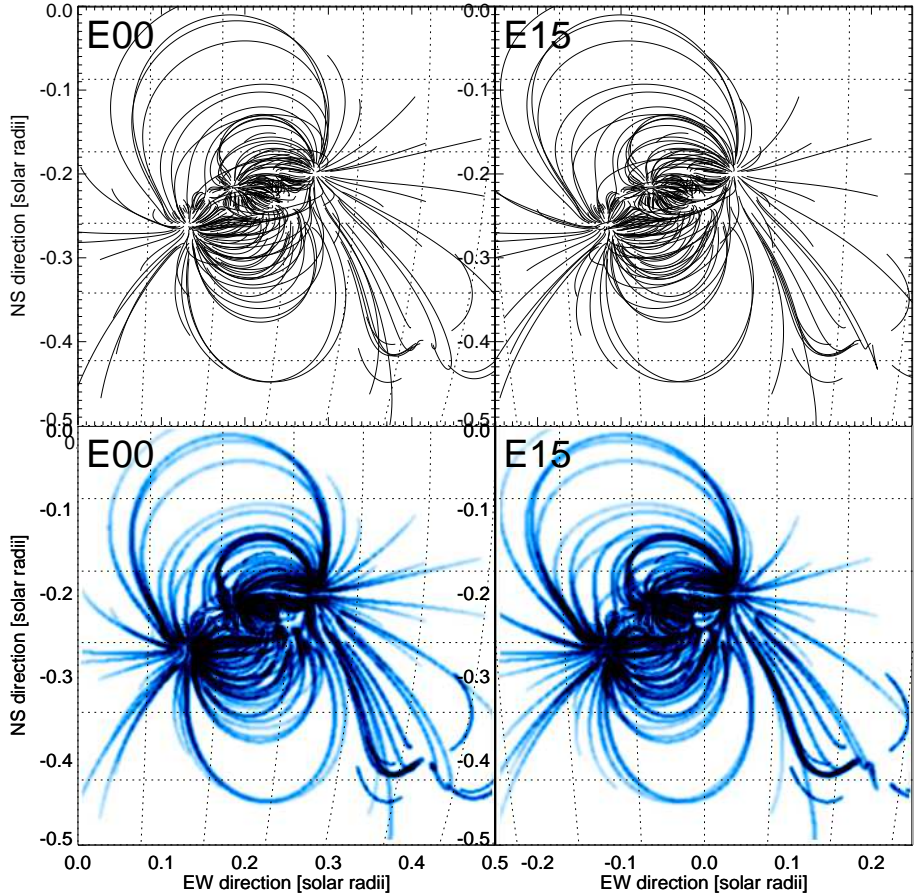}}
\caption{Simulated pair of stereoscopic images E15 and E00, seen from an
angle $15^\circ$ East from Earth (right bottom panel) and from Earth 
(left bottom panel). A heliographic coordinate grid is indicated with 
increments of $5^\circ$.  The simulated EUV images are composed from 261 
magnetic-field lines (top panels), generated from a potential-field 
extrapolation 
of a magnetogram of active region NOAA 11158 observed with HMI/SDO on 
15 February 2011, 01:14 UT. Only field lines with magnetic field strengths
of $|B_z| \ge 200$ G at the footpoints are displayed.}
\end{figure}

The barometric density is $n_e(h) \propto \exp(-h/\lambda)$, with the
scale height $\lambda = 50$ Mm $\times\ T_e$[MK], corresponding to a
temperature of $T_e=2$ MK, which is typical for structures that
are seen in the AIA 193\,\AA\ and 211\,\AA\ images. The intensity of
the image scales with the emission measure, \ie $F(x,y) \propto \int
EM\ {\rm dz} \propto n_e^2 \Delta z$, with $\Delta z$ the LOS-integrated
column depth. For sake of simplicity, we do not intend to simulate AIA
images at particular wavelengths, because the results of stereoscopic
simulations depend primarily on the geometry and signal-to-noise ratio
of the detected loops, which applies to any temperature or EUV
wavelength.

In order to create an adequately realistic EUV image, we convolve each
point of a field line with a Gaussian kernel that mimics typical loop
aspect ratios (of the loop width to the length) and gravitational
stratification. For the half width of a loop we choose the scaling of
$w(s) = w_0 \sqrt{n_s}$, where $n_s$ is the length of the loop in
pixels, and $w_0=1$ Mm is the minimum loop width. In order to mimic
the {\sl point spread function} (PSF) of the instrument, we
simulated loop widths [$w(s)$] with Gaussian kernels that are always
larger than the PSF, \ie $w(s) > w_{\rm psf} \approx$ two pixels.  

Simulated EUV flux maps of the optically thin plasma are rendered in
Figure 1 (bottom panels), similar to the method of Gary (1997). In the
later sections of this article we also simulate similar EUV maps with
different spacecraft-separation angles (section 6 and 7), with
different spatial resolutions (section 8), or with different magnetic
field models (section 10).

\section{	Automated Pattern Recognition  		}

Stereoscopy of coronal structures has been pioneered only by visual
tracing so far, for instance using {\it Skylab} images (Berton and Sakurai
1985) or STEREO/EUVI image pairs (Feng \etal 2007; Aschwanden et
al.~2008).  Manual tracing of coronal loops, however, is very
time-consuming and depends on human judgement, and thus is very
inefficient and subjective, preventing any frequent sampling or
real-time operation.  The availability of an automated 
pattern-recognition code is therefore a valuable element to accelerate 
progress in magnetic-field modeling of the
solar corona. In our context here, we aim for a ``blind stereoscopy
method'', rather than ``stereoscopy aided by visual guidance''.

Five experimental numerical codes for automated tracing of coronal
loops were compared in an initial study (Aschwanden \etal 1998). One
of them, the so-called oriented coronal curved loop tracing
(OCCULT-1) code was further developed and approached visual
perception (Aschwanden, 2010). A new advanced code (OCCULT-2) was
further optimized for curvi-linear tracing applied to {\sl Transition
Region and Coronal Explorer} (TRACE) data (Handy \etal 1999),
SDO/AIA data, Swedish Solar Telescope (SST) data, and
microscopic biophysics images (Aschwanden, DePontieu, and Katrukha
2013). Here, we use this OCCULT-2 code.  The automated pattern
recognition algorithm detects iteratively curvi-linear patterns with
large curvature radii, starting at a position with the highest flux,
propagating along the local ridge guided by the local curvature
radius, and it erases the signal of a traced loop segment from the
image before it starts with the next loop segment. The automated loop
tracings are carried out here independently in each image of a
stereoscopic pair (such as E00, E15 shown in Figure 2).

\begin{figure}
\centerline{\includegraphics[width=1.0\textwidth]{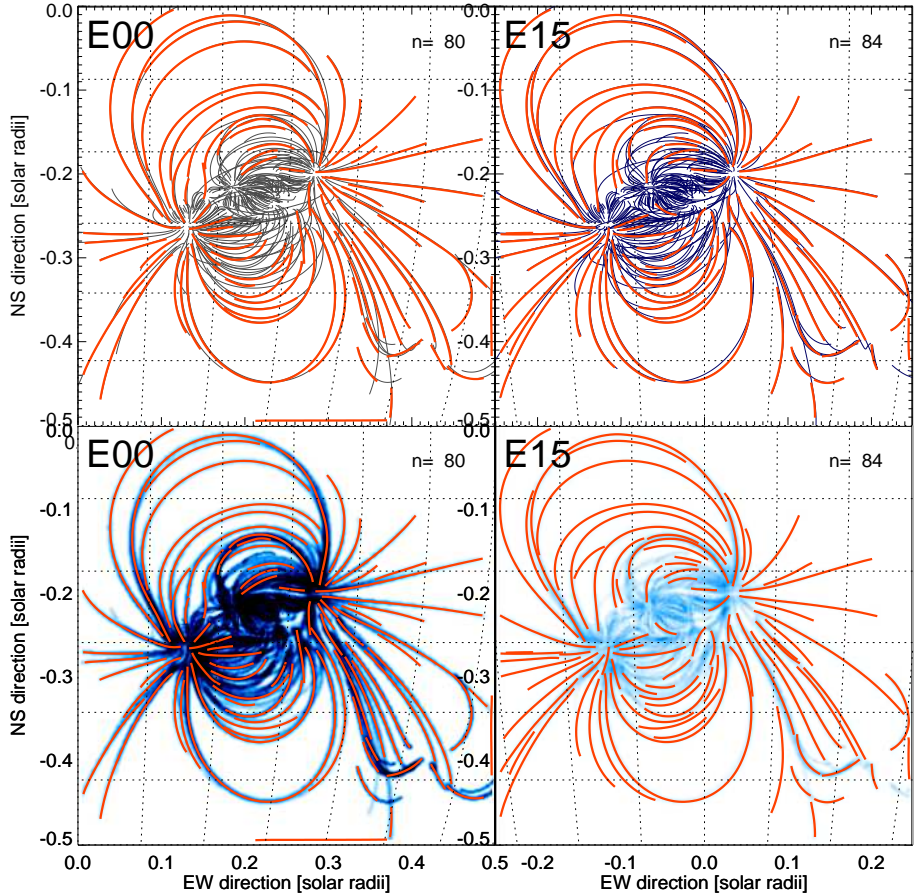}}
\caption{Automated loop tracing with the OCCULT-2 code (red curves),
superimposed on the simulated 261 magnetic-field lines (black curves 
in top panels) and the simulated EUV images (bottom panels).
A total of 80 (or 84) loop segments are detected at E00 (or E15) 
above a flux threshold of 1\% and with a minimum length of $l_{\rm min} 
\ge 20$ pixels.}
\end{figure}

For an example, we show the automated tracing of a pair of simulated 
stereoscopic images in Figure 2, which was simulated using 261 magnetic
field lines. The OCCULT-2 code detects from the EUV images a total of
80 loop segments with a length of $l_{min} \ge 20$ pixels in image E00,
and 84 loop segments in image E15.
The parameters can be adjusted depending on the type of data or desired 
pattern. What is particular about the simulated stereoscopic images 
here (Figure 1) is freedom from noise, in contrast to observed EUV images.
Noise-free images allow for more sensitive detection of faint structures
and are less prone to mis-guided detections in faint structures that are
comparable with the ambient noise level. We simulate noise-free images here in
order to study the performance of automated stereoscopy under ideal 
conditions, but will add data noise later to study the stereoscopic
behavior under more realistic conditions.

For our application here we chose the following parameter settings for
the code OCCULT-2: a pixel size of $1.5\arcsec$ (or 1.0\,Mm), a
highpass filter of $nsm_1=1$ pixel, a minimum curvature radius of
$r_{\rm min}=l_{\rm min}$ pixels, a minimum loop segment length of
$l_{\rm min}=20$ pixels, an image base level of $thr_1=0$, a threshold
level of $thr_2=0.01$ (in units of the highpass-filtered flux
maximum), no gap ($n_{\rm gap}=0$) along a traced structure, and a maximum
loop number of $n_{\rm max}=200$ loop structures per image. These settings
yield a near complete detection of unconfused loop segments, down to
the faintest structures seen visually (Figure 2). Since structures
are generally seen down to the spatial resolution of the instrument,
the same settings in units of pixels are recommended also for an
instrument with a different spatial resolution (although it corresponds
to a different absolute scale of the pixel size.) The most challenging
part is the crowded central core of the active region, where multiple
loops overlap and cross each other. A correct disentangling of loop
structures in such nested areas can probably only be achieved by
forward-fitting of multiple loop geometries, rather than by iterative
loop tracing. 

\section{	Automated Stereoscopic Pairing  		}

The second major task of the autonomous stereoscopy procedure is the pairing 
of corresponding loops, namely the correct association of loop segment $i$
in image A with the stereoscopic counterpart of loop segment $j$ in
the stereoscopic image B. This problem of ``stereoscopic correspondence''
or ``stereoscopic pairing ambiguity'' has never been systematically
investigated, and thus we explore it here to some degree to enable 
automated stereoscopy.

Stereoscopy is generally accomplished by transforming a stereoscopic
image pair into an epipolar coordinate system (Inhester 2006), which
generally requires a rotation and rescaling of each image (if the
images are taken with different image scales or at different distances
from the Sun). The epipolar plane is defined by three points: the Sun
center, and the positions of two stereoscopic vantage points, which are 
given by the spacecraft locations A and B for the solar STEREO
mission. In such an epipolar coordinate system, an image A is taken in
the $[x,y]$-plane, the epipolar rotation axis is in the $y$-direction,
which warrants that the stereoscopic parallax causes a rotational
shift in the $x$-direction only, while the $y$-coordinate remains
unchanged. Consequently, any structure with coordinate $[x(s),y(s)]$
in image A, where $s$ is a loop-length coordinate, has the coordinates
$[x(s) + \Delta x(\alpha_s) + \Delta x(h), y(s)]$ in image B,
where $\Delta x(\alpha_s)$ is the rotational shift of the
coordinate system due to the spacecraft-separation angle
$\alpha_s$, and $\Delta x[h]$ is the parallax that depends on the
altitude $h$ above the solar surface. The range of possible altitudes,
$0 \le h \le h_{\rm max}$, defines the solution space in image B, where
a corresponding loop can be located. In the ideal case, a loop is
detected over its entire length in both images, and appears isolated
within the search area. The search area for such a loop with
coordinates $[x_1(s), y_1(s)]$ in image A, is bound by $[x_2(s),
y_2(s)]$ with $y_1(s)=y_2(s)$ in the $y$-direction and $x_1(s) \le
|x_2(s)-\Delta x(\alpha_s)| \le \Delta x(h_{\rm max})$ in
the x-direction. If there is only one loop segment in this search area in
image B, the correspondence is unique and a stereoscopic triangulation
can directly be calculated. In reality, however, there are often
multiple loops in the search area and we have to develop a strategy to
find the most likely stereoscopically correct correspondence.

\begin{figure}
\centerline{\includegraphics[width=1.0\textwidth]{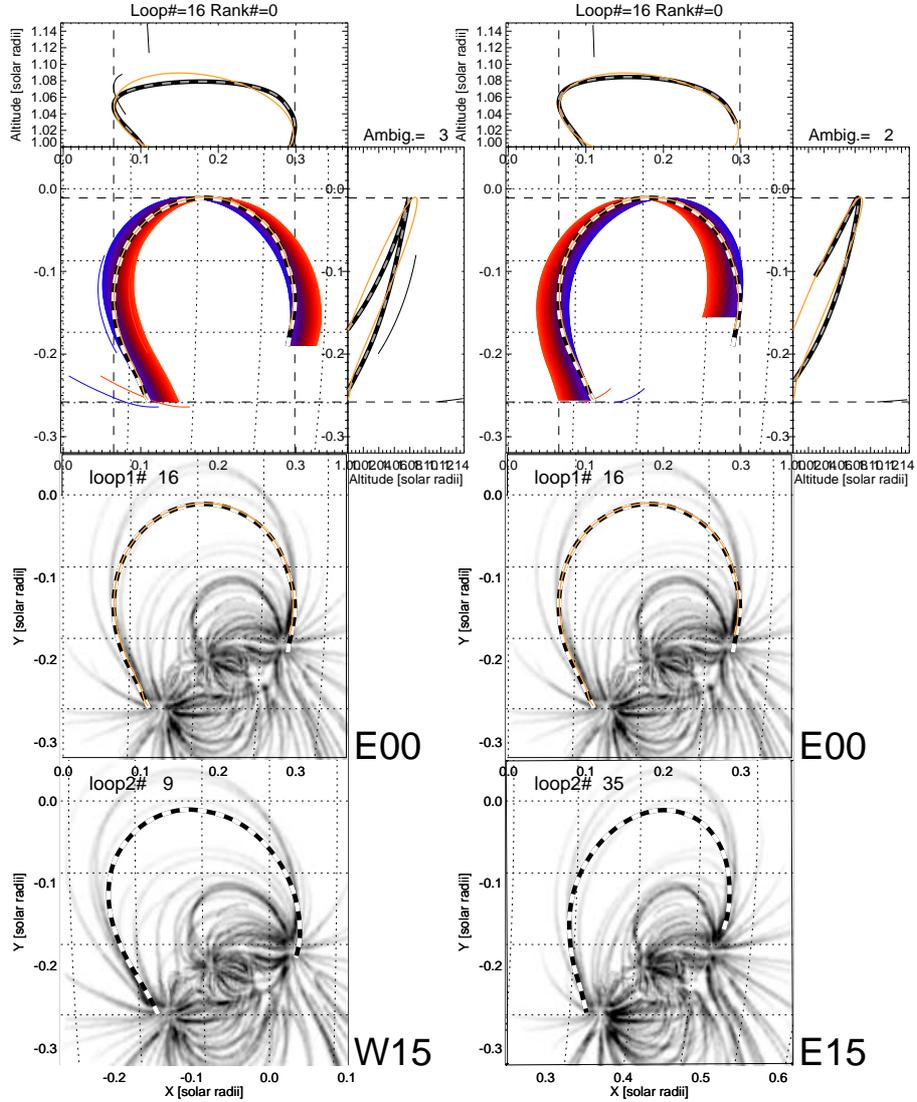}}
\caption{The largest loop traced in image E00 (second panel from bottom) 
is \#16 (black/white dashed linestyle), which corresponds to loop \#9 
in image W15 (bottom left panel), or to loop \#35 in image E15 (bottom
right panel). The projection of loops rotated at a photospheric level $h=0$ 
is indicated with blue curves, and rotated at a (maximum) coronal height of 
$h=0.15 {\rm R}_\odot$ with red curves. 
The side views (top panels) indicate the stereoscopically triangulated
altitudes of candidate loops in the solution space. The correct solution 
known from the simulated magnetic field is indicated with a field line 
drawn in orange color.}
\end{figure}

We illustrate the ``stereoscopic correspondence problem'' in Figure 3,
where we show stereoscopy between a spacecraft W15 and 
Earth view (E00) (Figure 3 left panels), as well as stereoscopy between
a spacecraft E15 and Earth view (E00) (Figure 3 right panels).
A total of $n=80$ loops were detected in image (E00), while 
$84$ loop segments were detected in image (E15), whereof
$n_C=51$ segments have an overlapping $y$-range in both images.
The tracing of the largest loop in image E00, which has the
number \#16 when sorted by length, is outlined (black/white
dashed linestyle) in Figure 3 in all panels. Rotating a loop 
structure from image E15 onto the view of E00, using two fixed 
distances from Sun center (\ie with a minimum altitude $h=0$ 
and a maximum altitude of $h_{\rm max}=0.15 {\rm R}_\odot$), we find 
a solution space in E00 for each structure detected in E15.
In this case we find two loop segments that overlap with 
the traced loop \#16 in image E00, so there is a two-fold ambiguity 
which loop should be stereoscopically triangulated. The boundaries 
of the solution space of loop \#35 from image E15 is indicated 
in the image E00 with a red--blue zone, where blue corresponds 
to a minimum altitude of $h=0$ and red to a maximum altitude of 
$h=h_{\rm max}=0.15$ solar radii. Which is the correct correspondence? 
Since the probability of a true correspondence increases with 
the length of the coincident segment, we use the criterion of 
maximum length, which indeed corresponds to the correct solution 
(indicated with an orange line in the top panels),
known from the simulated magnetic-field lines. 

How large is the ambiguity of pairing stereoscopic loop segments?
We count the number of loop segments in E15 to each loop segment 
of E00 that intersects with the stereoscopic solution space,
bound by an overlapping y-range and altitude range $h=[0, h_{\rm max}]$.
We find that most loops have an ambiguous stereoscopic correspondence, 
within a range of 1--10 possible correspondences, or a
statistical mean of $n_{\rm amb}=3.2\pm2.3$. In the example shown in Figure 3
(right panel), there are two ambiguous loop segments in image E15 that could 
potentially correspond to loop \#16 in image E00 within the 
altitude range used. The degree of ambiguity generally depends on the 
specified altitude range, which is chosen to be $h_{\rm max}=0.15$ 
solar radii here, and is expected to linearly increase with larger 
altitude ranges. One strategy to reduce the number of ambiguities 
is to eliminate those that have already been used previously in the 
iterative stereoscopic pairing.  A further strategy to avoid false 
stereoscopic pairings is to start with those that have the 
longest loop segments, where the least ambiguity occurs. 
Proceeding to smaller and smaller loop segments, the number of 
ambiguities then decreases systematically. 

Based on these considerations, we implement the following steps
in the (blind) stereoscopy code as a strategy to optimize the 
stereoscopic pairing procedure.

\begin{description}
\item{(i) All $n_A$ detected loops in A are sorted by their length.}
\item{(ii) All $n_B$ detected loops in B are sorted by their length.}
\item{(iii) For each sorted loop $i_{\rm A}=1,...,n_{\rm A}$ we determine which
        of the sorted loops $i_{\rm B}=1,...,n_{\rm B}$ overlap with the
	solution space of loop $i_{\rm A}$ within an altitude range
	of $h=[0,h_{\rm max}]$.}
\item{(iv) The longest segment in B that overlaps with the altitude
	range of loop $i_{\rm A}$ is selected as the stereoscopic
	counterpart $i_{\rm B,sel}$.}
\item{(v) A loop $i_{\rm B,sel}$ in B that has already previously been 
	paired with a loop $i_{\rm A}$, is excluded for pairing 
	in the next iterative pairing step.}
\end{description}

An alternative approach to solve the loop correspondence problem
is the so-called ``magnetic stereoscopy method'' (Wiegelmann \etal 2006; 
Feng \etal 2007), where an extrapolated linear force-free (LFFF) magnetic 
field is used to identify corresponding loops. A possible advantage of 
this method is that it reduces the solution space of corresponding loop 
locations more efficiently than our empirical stereoscopic pairing
method described above, especially for large spacecraft-separation
angles. However, a disadvantage of this method is that the initially 
chosen magnetic field (LFFF) model introduces a bias that favors 
solutions close to the initial (LFFF) model and may even prevent the 
convergence towards a nonlinear force-free field (NLFFF) solution.
However, the best method may be an iterative approach, where
stereoscopic loop pairing and magnetic field modeling is performed
in alternating steps, starting from an initial potential field model, a
and ideally ending at a best-fitting final non-potential field model.

\section{	3D-Triangulation of Loops  		}

The third step in the blind-stereoscopy procedure consists of the
triangulation of loop points. This is the easiest part of the 
stereoscopy procedure, because it is a uniquely defined 
mathematical geometry problem, after we have identified the
correct corresponding loop counterparts in both images A and B
in an epipolar coordinate system. Specific triangulation formula
are given in a number of previous studies (\eg Berton and
Sakurai 1985; Inhester 2006; Aschwanden \etal 2008).
Geometric parameters of the heliographic coordinate system
of an image are generally specified in the FITS headers
of the image data files (Thompson and Wei 2010).

Here, we derive the analytical relationships in their simplest 
form for a pair of two images A and B that have been already
rotated and scaled into an epipolar coordinate system, which
contains only four variables for every loop point $s$: 
$(x_{\rm A}, y)$ are the coordinates of a loop point in image A with
respect to the Sun center, $(x_{\rm B}, y)$ are the coordinates 
of the corresponding loop point in image B, where the $y$-coordinate
is identical in an epipolar coordinate system $(y=y_{\rm A}=y_{\rm B})$, and
the stereoscopic spacecraft-separation angle [$\alpha_s$],
measured from Sun center in the epipolar plane. Scaling the
distances in units of solar radii, the distance from Sun center
is $r = 1 + h$, where $h$ is the altitude above the solar surface.
The distance of point $(x_{\rm A}, y)$ from the solar (epipolar) axis 
is then
\begin{equation}
	\rho = \cos{b} \ (1 + h) \ ,
\end{equation}
where $(l,b)$ are the heliographic longitude and latitude 
in a Stonyhurst grid (with $l=0$ and $b=0$ at solar disk center).
The cartesian coordinates $x_{\rm A}, y_{\rm A}$, $x_{\rm B}$ are then related to
the heliographic coordinates $l_{\rm A}, l_{\rm B}, b$ by the following 
relationships,
\begin{equation}
	x_A = \rho \ \sin{(l_A)}  \ ,
\end{equation}
\begin{equation}
	x_B = \rho \ \sin{(l_B)}  \ ,
\end{equation}
\begin{equation}
	l_B = l_A + \alpha_s \ ,
\end{equation}
\begin{equation}
	y = \sin{(b)} \ (1 + h) \ .
\end{equation}
Now we can substitute and eliminate the observables $(l_{\rm A}, l_{\rm B}, y)$ 
and obtain the relationships for the variables $(\rho, b, h$),
\begin{equation}
	\rho = \left[ x_A^2 + \left( {x_B - x_A \ \cos{\alpha_s} 
	\over \sin{\alpha_s}} \right) \right]^{1/2} \ ,
\end{equation}
\begin{equation}
	b = \arctan{ \left( {y \over \rho} \right) } \ ,
\end{equation}
\begin{equation}
	h = {\rho \over \cos{b} } - 1  \ .
\end{equation}
Another simple method is to rotate the coordinate of a location
$(x_{\rm B}, y)$ from image B by the spacecraft angle $\alpha_s$ into
the coordinate system of image A using two different altitudes [$h_1$
and $h_2$], so that the correct altitude $h$ can be interpolated at
the matching position $x_B^{rot}=x_{\rm A}$. We used both methods in 
order to validate our triangulation code.

\begin{figure}
\centerline{\includegraphics[width=1.0\textwidth]{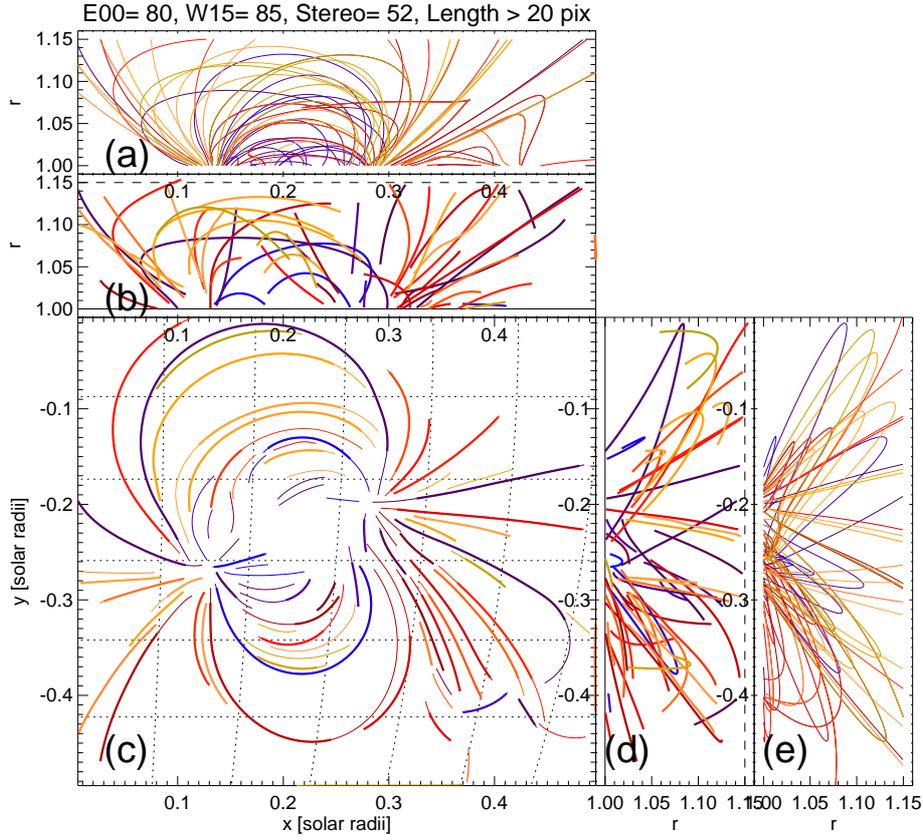}}
\caption{Automatically traced loops in image E00 (thin colored curves)
and loop segments with stereoscopic correspondences in E00 and E15 
(thick colored curves) are shown in the $x-y$ plane (panel c). 
The stereoscopically
triangulated solutions are shown in three projections, in the $x-r$ plane 
(panel b), and in the $r-y$ plane (panel d). The theoretical solutions
known from the magnetic field simulations are shown in the panels
a and e.}
\end{figure}

A result of stereoscopic triangulation of loops measured in the images
E00 and E15 is shown in Figure 4, along with their projections into 
orthogonal planes.
The 80 loop segments that were automatically traced in image
E00 are displayed in Figure 4c (thin solid curves), while
those segments for which a corresponding match in E15 was found are
indicated with thick solid curves. The triangulated heights are shown
as a function of the x-coordinate, $r(x)=1+h(x)$ (Figure 4b),
and as a function of the y-coordinate, $r(y)=1+h(y)$ (Figure 4d).
The closest matching 80 magnetic field lines that match the loops
traced in image E00 are indicated also (Figure 4a and 4e).
The display in Figure 4 demonstrates a good
match between the theoretical magnetic field lines and the 
stereoscopically triangulated loops.

The numerical accuracy of stereoscopic triangulation depends somewhat
on the spatial direction of the coronal loop or magnetic field line.
In the epipolar coordinate system, the stereoscopic parallax occurs 
in the $x$-direction, which yields the most accurate measurement if a loop
or field line is oriented in the y-direction, \ie in the North-South
direction. If the loop is oriented in the x-direction, there is a
singularity in the altitude inversion, because the parallax direction
coincides with the loop direction, and the correspondence of a loop
segment in a pair of two stereoscopic images is mathematically
ill-defined, which prohibits stereoscopic triangulation at this location. 
This singularity, which we may call the ``epipolar degeneracy'', can 
affect the accuracy of stereoscopic triangulation for a range of angles
where the tilt-angle 
$\tan(|\vartheta(s)|)=|y(s_{i+1})-y(s_i)| / |x(s_{i+1})-x(s_i)|$ 
along a loop coordinate $s$ has
a small value, due to the finite spatial resolution and directional
tracing errors (Aschwanden \etal 2008, 2012b). In
order to overcome this epipolar degeneracy problem, we apply
stereoscopic triangulation only at loop locations [$s$] where
the loop direction is larger than a critical value, i.e., 
$|\tan{\vartheta(s)}| \ge 0.1$, and apply a low-order polynomial 
interpolation for the coordinate $z(s)$ in those gaps. 

\section{	Dual Versus Triple Spacecraft Stereoscopy 		}

The minimum option for solar stereoscopy is two perspectives, but one
may opt for a third for a number of reasons. First of all, should one
instrument fail, one has still a fall-back option with two spacecraft
that can essentially accomplish the stereoscopy task. Secondly,
stereoscopy with two perspectives often confronts us with the problem
of ambiguous correspondence. Which loop structure from perspective A
has to be triangulated with what loop structure from perspective B?
The combination of three perspectives virtually eliminates the
stereoscopic-ambiguity problem.

\begin{figure}
\centerline{\includegraphics[width=1.0\textwidth]{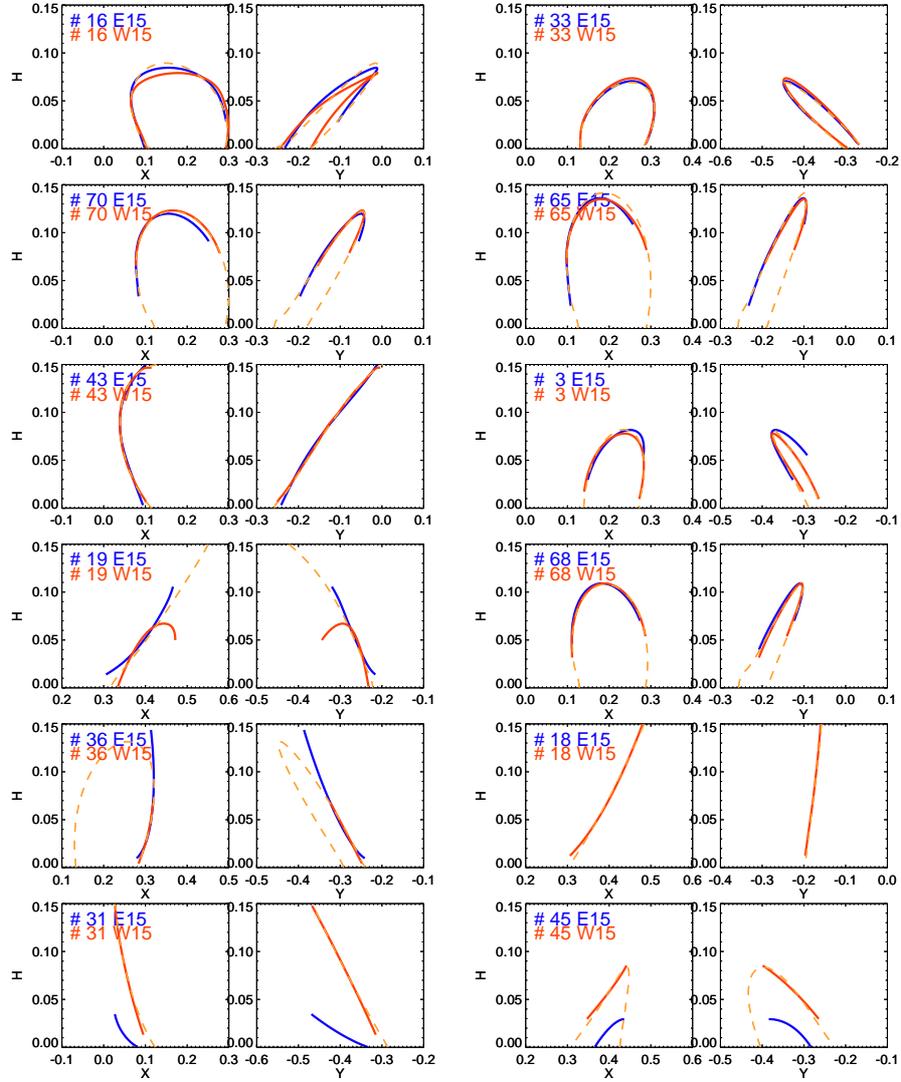}}
\caption{Comparison of solutions for altitudes as a function of the 
$x$-coordinate, $h(x)$, and the $y$-coordinate, $y(h)$, from the stereoscopic
triangulation between the images E00 and E15 (blue curves), and the images
E00 and W15 (red curves). The theoretical solutions based on the closest 
simulated magnetic-field lines are indicated in orange. Inconsistent 
solutions occur for four loops (\#18, 19, 31, 45) out of the 12 cases, but are 
correct for the spacecraft image W15 (red curves) in all cases but \#19.}
\end{figure}

We demonstrate the bootstrapping effect of stereoscopic triangulation
as it could be achieved with three spacecraft in the example shown in
Figure 5. The view of the active region from the three spacecraft E00,
E15, and W15 is depicted in Figure 3, and stereoscopy of loop \#16 from
the image pair E00 and E15 (Figure 3 right panels), and the pair E00 and
W15 (Figure 3 left panels) yields a self-consistent solution that is
close to the theoretical values (orange curves in Figure 3). In Figure 5
(top left panels) we show the orthogonal projections $h(x)$ and $h(y)$
for the same loop \#16, found from the spacecraft pair E15+E00
(Figure 5, blue curves), and from the spacecraft pair W15+E00 (Figure 5,
red curves), along with the theoretical solution (Figure 5, orange
curves in dashed linestyle), which all agree within $\Delta h \lapprox
0.01$ solar radii. In the same representation we show the same
information for the 12 longest detected loops in Figure 5. From these 12
stereoscopic triangulations we see a consistent solution of both
spacecraft pairs with the theoretical model field lines in eight cases,
while the results from E15+E00 (Figure 5, blue curves) fail for the four
loops \#18, 19, 31, and 45. Nevertheless, the results from W15+E00 are
correct in 11 out of 12 cases. For this particular example, which may
be typical for many other observations, we can say that stereoscopic
triangulation with three spacecraft is successful in $\approx 90\%$
(11 out of 12 cases), while triangulation with two spacecraft can have
a reduced success rate in the range of 75\%--90\%, depending on the
position of the spacecraft.  A generalization of our two-spacecraft
stereoscopy code to a triple-spacecraft configuration could easily 
be implemented,
based on the relative overlap range of the $y(s)$-values of the
automatically traced loop segments in each of the three spacecraft
images, and the highest probability or correct stereoscopic
correspondence based on the maximum lengths of the paired loop
segments among the three images from different vantage points.

\section{		Spacecraft Separation Angle 		}

What is the optimum spacecraft-separation angle for stereoscopy?
In a previous study with STEREO/EUVI data, it was demonstrated
that stereoscopic triangulation is in principle possible from small 
($6^\circ$) to large ($170^\circ$) angles, based on a small sample 
of visually traced loops (Aschwanden \etal 2012b). Combining the
accuracy of altitude triangulation with the stereoscopic correspondence
ambiguity, it was estimated that a spacecraft-separation angle of
$\alpha_s=22^\circ-125^\circ$ is most favorable for stereoscopy,
using an instrument with the spatial resolution $2.6\arcsec$
(with pixel size of $1.6\arcsec$) such as STEREO/EUVI.

Here, we simulate data for a spacecraft-separation angle of 
$\alpha_s=1^\circ$ to $90^\circ$ in both the eastern and western 
direction, and perform stereoscopic triangulation between a spacecraft 
at angle $\alpha_s$ (W90,...,W01, E01, ...E90) and the spacecraft 
at Earth view (E00). For simplicity we label the positions at angles
(W90,...,W01, E01, ...E90) with ``spacecraft A'', and the position at 
Earth view with ``spacecraft E''. 

\begin{figure}
\centerline{\includegraphics[width=1.0\textwidth]{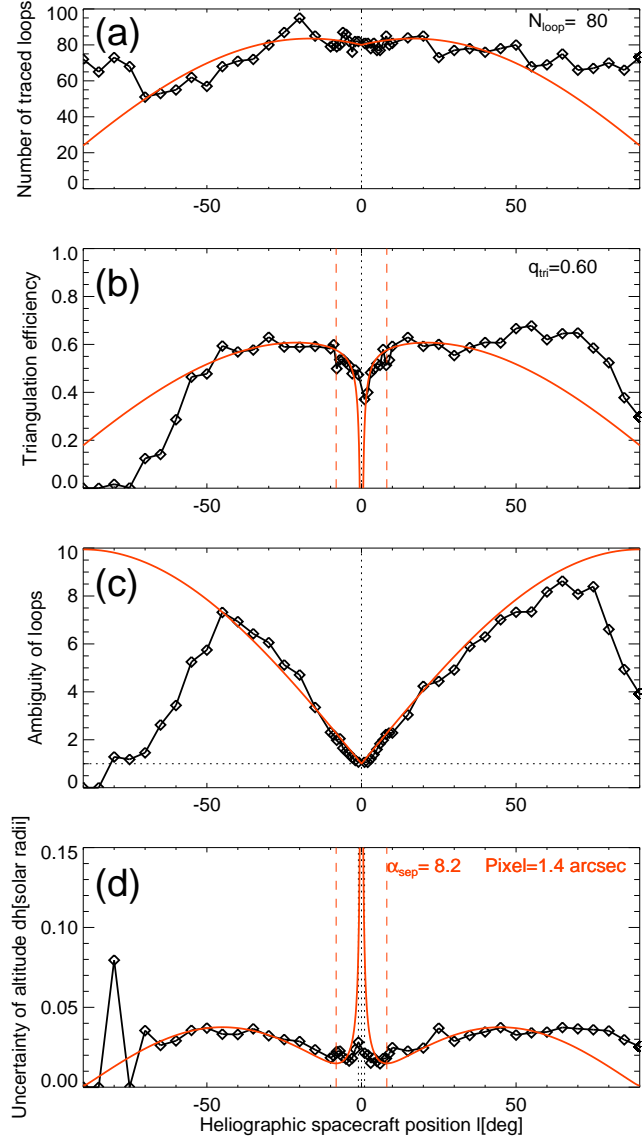}}
\caption{(a) Statistics of the number of automatically traced loops (top panel),
(b) the number of triangulation points (second panel), (c) the
mean number of ambiguous loops per triangulated loop, given by the mean 
and standard deviation (third panel), and (d) the median accuracy of altitude
measurements $dh$ in units of solar radii, as a function of the spacecraft
heliographic position (or spacecraft-separation angle from Earth).
The colored curves correspond to the theoretical model described in the text.
The spatial resolution corresponds to a pixel size of 
$\Delta x=1.4\arcsec$.}
\end{figure}

For each of the spacecraft positions we measure the number of
automatically traced loops (Figure 6a), which has a maximum value of
$N_0 = 80$ (in image E) and tends to decrease as the region is seen
closer to the solar limb (in the images A).  If the distribution of
loops were homogeneous and isotropic in an active region, the view would be
rotation-invariant and the number of detected loops should be constant
as a function of the aspect angle. Therefore, the observed slight
decrease of detected loops towards the limb indicates a larger
horizontal than vertical extent of the active region. If we assume a
homogeneous density of loops in a box with a horizontal length [$\Delta
x$] and height [$h_{\rm max}$], the projected length of the box as a function
of the rotation angle (or longitude) $\alpha_s$, we expect that
the number of detected loops is roughly proportional to the projected
length of the box, for which we then expect a center-to-limb variation
of
\begin{equation}
	N_{loop}(\alpha_s) \approx N_0  
	{\Delta x \cos{ (\alpha_s)} + h_{max} \sin{ (\alpha_s)} 
	\over {\Delta x} } \ ,
\end{equation} 
where $N_0$ is the maximum number of detected loops, $\Delta x$ is the 
East--West extension of the active region, and $h_{max}$ is the maximum 
altitude. We overplot such a function in Figure 6a, using $\Delta x=0.5 
{\rm R}_\odot$ based on the chosen field-of-view, 
and $h_{max}\approx 0.15 {\rm R}_\odot$, 
which approximately reproduces the decrease of detected loops towards
the limb. The AR is located at $12^\circ$ West, causing an obscuration
by the limb at $-78^\circ$ and the number of detected loops to go to
zero beyond the limb.

Then we measure the triangulation efficiency $q_{\rm tri}$ in suitable 
loop segments (Figure 6b). This number is defined by the ratio 
$q_{\rm tri}=N_{\rm tri}/N_{\rm all}$ of the number of loop positions $N_{\rm tri}$
where a valid stereoscopic triangulation could be executed, 
normalized to the total number $N_{\rm all}$ of all possible loop positions.
The requirement for a stereoscopic triangulation of a loop position
is a valid stereoscopic correspondence between two spacecraft A and E,
which is an identical $y$-position in both spacecraft A and E, and a 
valid altitude range of $0 \le h \le h_{\rm max}$ in the stereoscopic 
triangulation. For this number we find a typical value of 
$q_{tri}\approx 0.5$, mostly caused by incomplete loop tracing or 
erroneous stereoscopic correspondence identifications.
As we can see in Figure 6b, the number of triangulated
loop positions $q_{\rm tri}(\alpha_s)$ follows a similar function 
as the number of detected loops (Equation 9), with a drop-off at both 
the eastern and western limb. Thus the efficiency of stereoscopy 
is warranted in a broad range
of $|\alpha_s-l_0| \lapprox 60^\circ$ between a spacecraft position A
and E, where $l_0 \approx 12^\circ$ is the longitude of the active region.

\begin{figure}
\centerline{\includegraphics[width=1.0\textwidth]{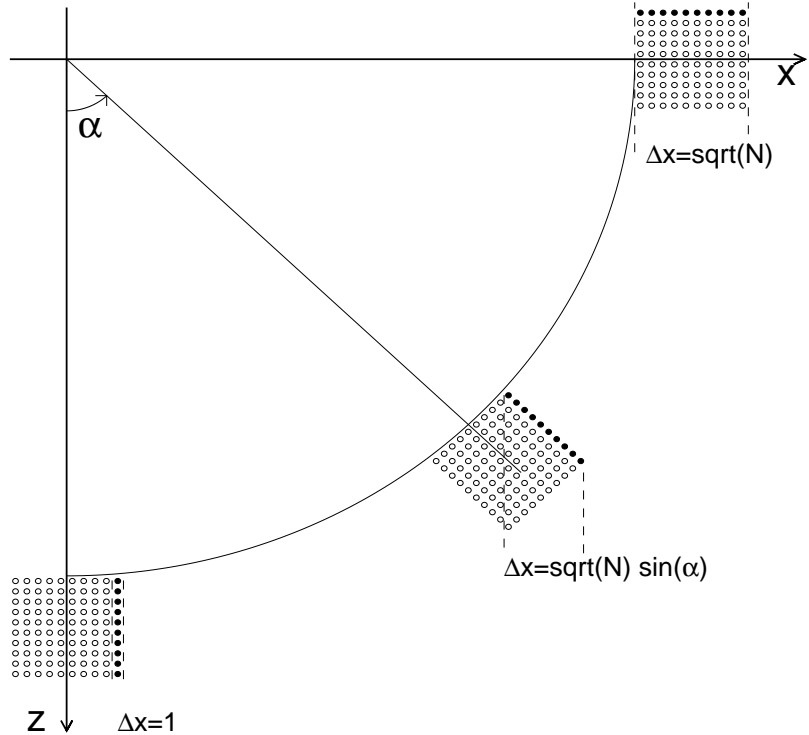}}
\caption{The number of ambiguous loops in
stereoscopic correlation as a function of the spacecraft-separation
angle [$\alpha_s$] is shown. If spacecraft A located at $\alpha=0^\circ$ detects
one single loop within a spatial width of $\Delta x=1$, a spacecraft
B at position $\alpha=90^\circ$ detects a number of $N_{\rm amb}(\alpha)
=\sqrt{N}$ loops that all have the same projection after they have been rotated
by an angle of $-\alpha$ to the viewpoint of spacecraft A in the 
$z$-direction. For an intermediate position B at $\alpha$, the number
of ambiguous loops scales as $N_{\rm amb}(\alpha)=\sqrt{N} \sin(\alpha)$.}
\end{figure}

The number of ambiguous loops may also play a significant role in the 
evaluation of the optimum spacecraft angle for stereoscopy. We performed
automated stereoscopy between a near-Earth spacecraft E and a
spacecraft A at any position from the most western viewpoint at 
$\alpha_s=-90^\circ$ to the most eastern viewpoint
$\alpha_s=+90^\circ$, in increments of $5^\circ$ for the whole
range, and in increments of $1^\circ$ in the small-angle range of
$-10^\circ \le \alpha_s \le +10^\circ$. While the automated code
detected $N_{\rm loop}=80$ loop structures in image E, an average number of
$1 \le n_{\rm loop,A} \lapprox 10$ ambiguous loops were detected in image A, 
where ``ambiguous'' means the number of candidate loops in image A
that have a valid altitude range $0 \le h \le h_{\rm max}=0.15$ in the
stereoscopic triangulation. The variation of the number of ambiguous
loops is shown in Figure 6c, which reveals a minimum of one single
loop at $\alpha_s \approx 0^\circ$,
while the ambiguity seems to increase up to angles of 
$\alpha_s \approx 50^\circ-70^\circ$. We can model the number
of ambiguities by assuming a uniform distribution of North-South
oriented loops in a cube, which is most favorable for stereoscopy (Figure 7).
If we have $n_x$ loops in $x$-direction and $n_h$ loops in $h$-direction,
the total number of loops is $N_{\rm loop}=n_x \times n_h$. For a quadratic
box with a maximum number $N_{\rm loop}$ of detected loops we have 
$n_x = n_h =\sqrt{N_{\rm loop}}$. The number of ambiguous loops scales then
with the projected length ($\Delta x$ in Figure 7) as a function of the
rotation angle (or spacecraft-separation angle $\alpha_s$).
If a loop were detected within a width of $\Delta x=1$ (Figure 7)  
for a small stereoscopic viewing angle, the projected width of all
stereoscopically corresponding loops is  $\Delta x \approx \sqrt{N}$ 
at the limb (Figure 7), and scales according to the sine-function
inbetween,
\begin{equation}
	n_{amb}(\alpha_s) = 1 + \sqrt{N_{loop}}
		\ \sin{ |\alpha_s| } \ ,
\end{equation}
which is overplotted on the measurements in Figure 6c. The predicted
number of ambiguous loops matches the numerically determined number
fairly accurate over the range of $-50^\circ \lapprox \alpha_s
\lapprox 80^\circ$.  The ambiguity factor is almost
symmetrical for a western and eastern separation angle, but exact
symmetry is not expected for an active region that has no symmetry
in its magnetic field and EUV brightness.

\begin{figure}
\centerline{\includegraphics[width=1.0\textwidth]{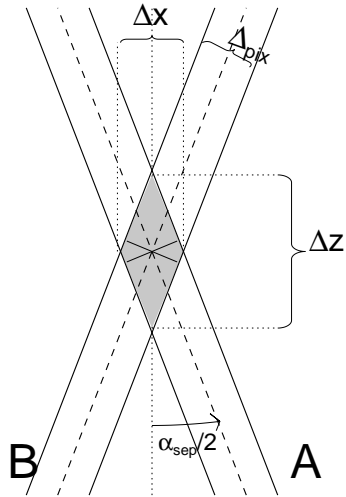}}
\caption{The error trapezoid of stereoscopic triangulation is shown
(grey area), given by the two lines of sight of the two observer
directions A and B, separated by an angle $\alpha_s$. The 
uncertainties [$\Delta x$] in the $x$-direction and in
the $z$-direction depend on the pixel width [$\Delta_{\rm pix}$] and
half aspect angle $\alpha_s/2$ (Aschwanden \etal 2012b).}
\end{figure}

The ultimate parameter that determines the optimum spacecraft-separation
angle for for dual-spacecraft stereoscopy is the accuracy of altitude
measurements, which depends not only on the ambiguity factor (for
stereoscopic correspondence) but also on the spatial resolution of
the images. If we make the spacecraft-separation angle smaller and
smaller, the horizontal parallax as a function of the altitude
becomes also smaller and will reach sub-pixel scale, where the 
stereoscopic information becomes unmeasurable, because there is a
singularity when the spacecraft-separation angles approach zero.
Theoretically, the error in the line-of-sight extent [$\Delta z$]
of a point source observed with a spatial resolution [$\Delta x$]
can be understood from the ``trapezoid relationship'' (Aschwanden 
\etal 2012b) shown in Figure 8, 
\begin{equation}
	\Delta h_{res} = {\Delta x / 2 \over \sin{(|\alpha_s|/2)} } \ ,
\end{equation}
which exhibits a singularity at spacecraft-separation angle $\alpha_s
=0$ in the number of stereoscopic triangulation points (Figure 6b) and
in the accuracy [$\Delta h$] (Figure 6d) of stereoscopically triangulated 
altitudes.  For larger spacecraft-separation angles, this
finite spatial resolution effect becomes negligible, while uncertainties
due to the ambiguity factor in stereoscopic correspondence dominate.
Thus, the center-to-limb variation of the accuracy of altitude 
measurements is expected to vary as a sine-function as the ambiguity
function does (Equation 10), where the maximum (positive or negative)
error corresponds to the half height range ($h_{\rm max}/2)$,
\begin{equation}
	\Delta z_{amb} = {h_{max} \over 2} \sin{(|\alpha_s|)} \ .	
\end{equation}
In order to obtain an error in altitude [$\Delta h$], we have to 
correct for the cosine-angle of the projection between the line-of-sight
$z$ and the altitude $h$,
\begin{equation}
	\Delta h_{amb} = \Delta z_{amb} \ \cos{|\alpha_s|} \ ,
\end{equation}
which yields the combined error (added in quadrature),
\begin{equation}
	\Delta h = \sqrt{ \Delta h_{res}^2 + \Delta h_{amb}^2 } = 
	\left[
	\left(
	{\Delta x / 2 ) \over \sin{(|\alpha_s|/2)} }
	\right)^2 +
	\left(
	{h_{max} \over 2} {\sin{\alpha_s} \cos{\alpha_s)}}
	\right)^2 
	\right]^{1/2}\ .
\end{equation}
This theoretical prediction is overplotted on the numerical
datapoints of the uncertainties for the stereoscopic altitude
measurements (Figure 6d, red smooth curve), which matches the data closely
in the range of $-60^\circ \le \alpha_s \le +60^\circ$.
An interesting consequence of this
model is that it predicts a highest accuracy at a spacecraft
separation angle of $\alpha_s \approx 8.2^\circ$. 
According to the simulations, the highest accuracy
is $\Delta h \approx 0.02 {\rm R}_\odot$ (or 14 Mm), 
does not deteriorate more than about a factor of two 
to $\Delta h \lapprox 0.04 {\rm R}_\odot$  (or 28 Mm) at larger
separation angles. 

\begin{figure}
\centerline{\includegraphics[width=1.0\textwidth]{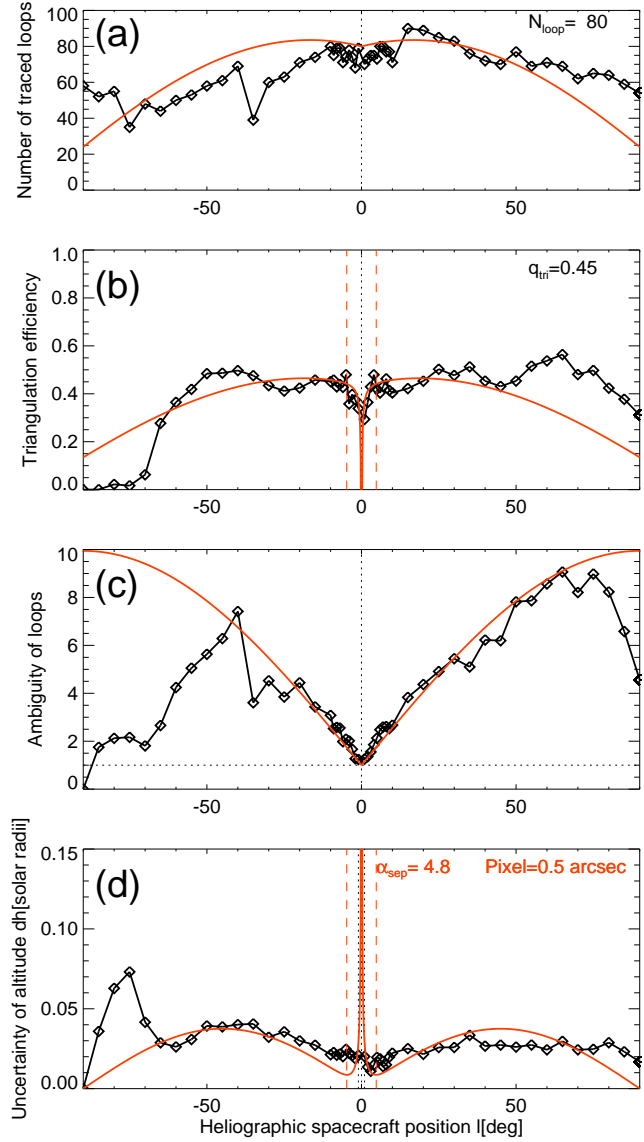}}
\caption{Same representation as Figure 6, but for a spatial resolution 
corresponding to a pixel size of $\Delta x=0.6\arcsec$):
(a) Statistics of the number of automatically traced loops (top panel),
(b) the number of triangulation points (second panel), (c) the
mean number of ambiguous loops per triangulated loop, given by the mean 
and standard deviation (third panel), and (d) the median accuracy of altitude
measurements $dh$ in units of solar radii, as a function of the spacecraft
heliographic position (or spacecraft-separation angle from Earth).}
\end{figure}

This is also the reason why a relatively
wide range of stereoscopic angles of $\alpha_s \approx
22^\circ-125^\circ$ was found to be usable for stereoscopy
according to an earlier study (Aschwanden \etal 2012b), 
a range that has been adopted for another proposed
future stereoscopic mission (Strugarek \etal 2015).
Note that the uncertainty of the ambiguity was estimated
differently in the previous study, assuming a dependence of
$\Delta h_{\rm amb} \propto 1/|\cos{(\alpha_s)}|$ based on
a theoretical argument (Equation 7 in Aschwanden \etal 2012b), 
while we obtain a probably more realistic dependence of
$\Delta h_{\rm amb} \propto \sin{(\alpha_s)} \cos{(\alpha_s)}$ 
(Equations 12 and 13) here, based on numerical simulations of 
stereoscopic triangulations. The large range of $\alpha_s 
= 22^\circ-125^\circ$ still specifies an angular range where
stereoscopy is feasible, if we tolerate a factor of two
in the uncertainty of altitude measurements (Figure 6d and 9d),
but the optimum angle of $4.6^\circ-10.7^\circ$ found here
(Table 1) yields the most precise measurements according to
our simulations.

\begin{table}
\caption{Dependence of optimum spacecraft-separation angle
[$\alpha_{\rm best}$] and the uncertainty of stereoscopic altitudes 
[$\Delta h$] on the spatial resolution [$\Delta x$] and altitude
range [$h_{\rm max}$] of the solution space.}
\begin{tabular}{lcccccc}
\hline
Instrument & Pixel & Spatial    & Pixel & Altitude & Spacecraft & Altitude   \\
	   & size  & resolution & size  & range    & separation & uncertainty\\
	   &       &            &       &          & angle      &            \\
	   & $\Delta x$ & PSF & PSF & $h_{max}$ & $\alpha_{best}$ & $\Delta h$\\
           & [arcsec] & [arcsec]& [Mm]  & [Mm]     & [deg]      & [Mm] \\
\hline
AIA		& 0.6 & 1.4 & 0.44 &  70  &  6.3$^\circ$ &   5.5    \\
EUVI		& 1.6 & 2.6 & 1.16 &  70  & 10.7$^\circ$ &   8.9    \\
\hline
AIA     	& 0.6 & 1.4 & 0.44 & 140  &  4.6$^\circ$ &   7.8    \\
EUVI		& 1.6 & 2.6 & 1.16 & 140  &  7.4$^\circ$ &  12.6    \\
\hline
\end{tabular}
\end{table}

\section{		Spatial Resolution  			}

The accuracy of stereoscopic altitude measurements apparently
depends on the spatial resolution $\Delta x$ of the instrument,
the spacecraft-separation angle $\alpha_s$, and the altitude
range $h_{\rm max}$ of the solution space, according to the 
relationship of Equation (14), and thus the best spacecraft-separation
angle depends on the same parameters [$\Delta x$] and [$h_{\rm max}$].
We determined the minimum of the function $\alpha_s(\Delta x,
h_{\rm max})$ numerically and tabulate the values for $\Delta x=
0.6\arcsec, 1.6\arcsec$ and
$h_{\rm max}=[0.1, 0.2] {\rm R}_\odot$ in Table 1. The coarsest
spatial resolution of $2.6\arcsec$ (with pixel size of $1.6\arcsec$)
corresponds to the EUVI/STEREO instruments, and $1.4\arcsec$ 
(with pixel size of $0.6\arcsec$) to the SDO/AIA instrument with 
{\it $4k\times 4k$} CCD cameras.

The values in
Table 1 show that the spacecraft-separation angle increases from
$\alpha_{\rm best}=6.3^\circ$ to $\alpha_{\rm best}=10.7^\circ$ for the
altitude range of $h_{\rm max}=0.1 {\rm R}_\odot$, while the accuracy of the
stereoscopically triangulated altitudes worsens steadily from
$\Delta h=5.5$ Mm to 8.9 Mm. Thus, the highest accuracy is clearly
achieved for the instrument with the highest spatial resolution, as
expected.  We can derive an approximate relationship for the optimum
spacecraft-separation angle $\alpha_{\rm best}$ by setting the uncertainty
due to the spatial resolution ($\Delta h_{\rm res}$; Equation 11) equal to the
uncertainty due to loop pairing ambiguities ($\Delta h_{\rm amb}$; Equation 12
and 13). For small spacecraft-separation angles we can then use the
approximations $\sin(\alpha_s) \approx \alpha_s$, and
$\cos(\alpha_s) \approx 1$, which yields the relationship
\begin{equation}
	 \alpha_{best} \approx \sqrt{ 2 \Delta x \over h_{\rm max} } 
	\ , \qquad [{\rm rad}],
\end{equation}
which tells us that the spacecraft-separation angle [$\alpha_s$]
scales with the square root of the spatial resolution [$\Delta x$].
Varying the altitude range $h_{\rm max}$ by a factor of two,
the stereoscopic accuracy worsens a factor of $\sqrt{2}$.
Thus, for a typical range of $h_{\rm max} \approx (0.1-0.2) {\rm R}_\odot$,
and for a pixel size $0.6\arcsec$ (\ie $\Delta x=0.000625$)
as used for the AIA instrument, the optimum spacecraft-separation angle 
is $\alpha_{\rm best}\approx 4.6^\circ-6.3^\circ$. 
We repeat the calculations, shown for a spatial resolution of 
$1.6\arcsec$ shown in Figure 6, for the AIA spatial resolution of
$0.6\arcsec$ in Figure 9. If we use the existing
STEREO/EUVI instruments with a spatial resolution of $2.6\arcsec$
(with pixel size of 1.6\arcsec, or $\Delta x=0.0017$), the optimum 
spacecraft-separation angle
would be in the range of $\alpha_s \approx 7.4^\circ-10.7^\circ$
(Table 1).

\section{		Spacecraft Position  			}

Considering the functional behavior of $\Delta h(\alpha_s)$ as
shown in Figures 6 and 9, we also expect a minimum of the stereoscopic error
at large angles of $\alpha_s=\pm90^\circ$. This solution
corresponds to an orthogonal view of an active region, where the
$x$ and $z$ position of a loop can be measured with maximum accuracy.
This large-angle configuration, however, has at least three major
disadvantages compared with the optimum small-angle configuration:
(i) Obscuration by the solar limb has a much higher probability that
an active region is not seen simultaneously by two spacecraft;
(ii) The stereoscopic correspondence ambiguity is much more severe
at large spacecraft angles than at small ones; and (iii) A large
spacecraft distance from Earth demands a much higher telemetry
power. For instance, a spacecraft-separation angle of $\alpha_s
=6^\circ$ implies a ten times smaller proximity to Earth than a
spacecraft position of $\alpha_s \approx 60^\circ$ at 
Lagrangian points L4 or L5, and thus enables a 100 times higher
telemetry rate, or a 100 times smaller telemetry power for the same
data rate.

\begin{figure}
\centerline{\includegraphics[width=1.0\textwidth]{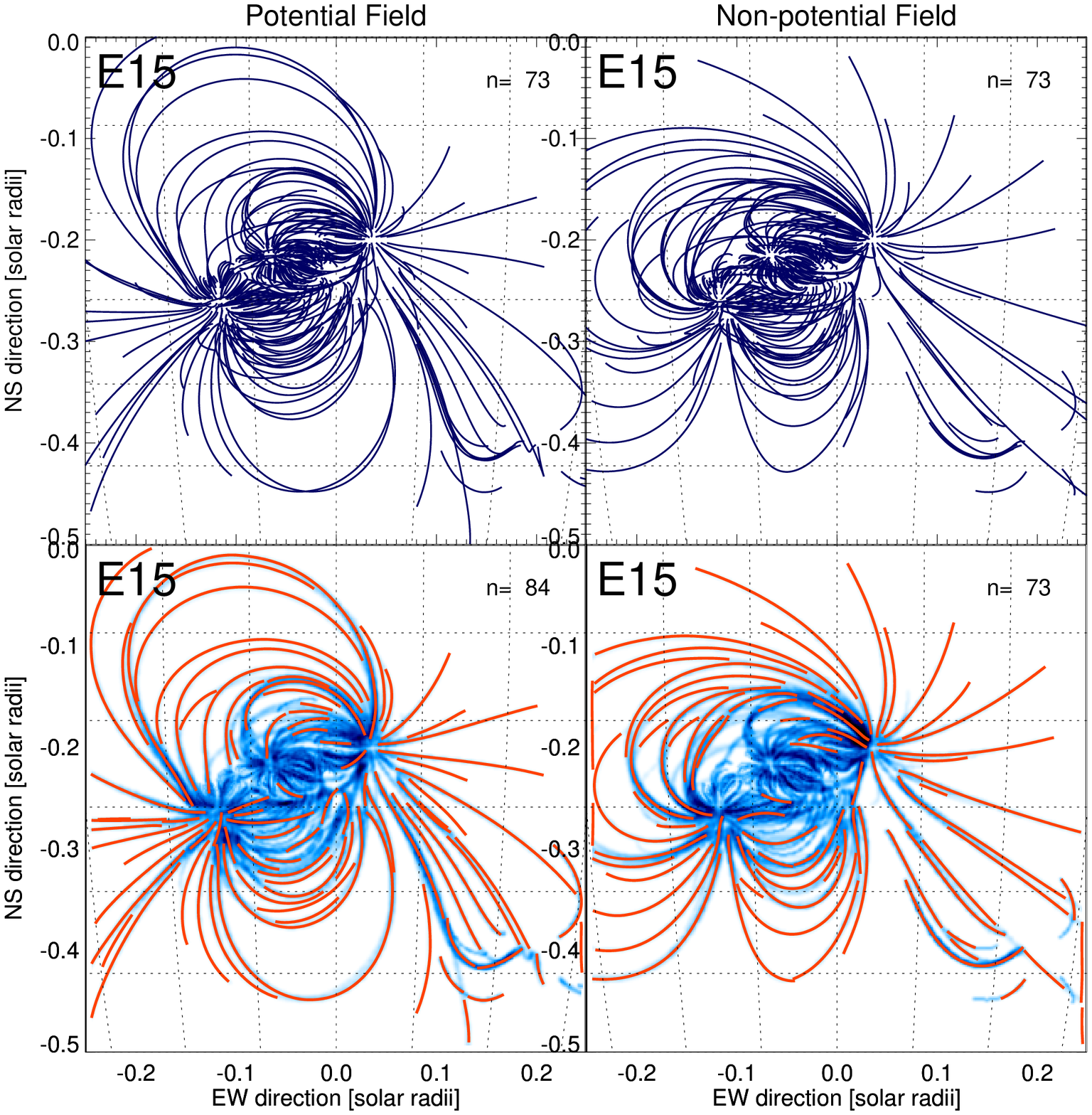}}
\caption{Automated loop tracing with the OCCULT-2 code (red curves),
superimposed on the simulated magnetic field lines (blue curves in 
top panels) and the simulated EUV images (bottom panels).
The left panels show a potential field, while the right panels show
a non-potential field. Note that the tracings of loops in the EUV
images (bottom) are significantly different for the potential and
non-potential fields, which proves that our automated method is
very sensitive to the degree of the magnetic non-potentiality.}
\end{figure}

\section{	Non-Potentiality of Magnetic Field }

The ultimate goal of a new stereoscopy mission is a reliable and
accurate method to measure the magnetic field of solar active regions
in order to monitor the evolution of non-potential fields and to
determine their free energy that can be released in large solar flares
and coronal mass ejections. The question arises then, whether the
proposed automated tools are sufficiently accurate for this task.
We may ask whether a stereoscopy method is sufficiently
sensitive to distinguish between a potential and non-potential
magnetic field, and how accurately can it measure the degree of
non-potentiality. For a qualitative demonstration we simulate the
magnetic field for a potential field, using 100 unipolar magnetic
charges to represent the field (Figure 10; top left), and simulate a
corresponding EUV image in the same way as described above (Figure 10;
bottom left), to which then our OCCULT-2 code is applied to trace the
loops in an automated way (red curves in Figure 10, bottom left).

Then we use the same model with 100 unipolar magnetic charges, but add
a vertical electric current to the strongest magnetic charge, which
has a magnetic-field strength of -2070 G and is located at
$(x,y,z)=(-0.13,-0.26,0.95)$ (see left-most sunspot in Figure 10) by
assigning a force-free parameter $\alpha$-parameter of $\alpha_1=2 \pi
N_{\rm twist}/L = 10^{-9}$ cm$^{-1}$. The method of parameterizing the
nonlinear force-free field is identical to that of the
parameterization used in the simulation of the magnetic-field data
described in section 3 (Aschwanden \etal 2012a).
Comparing this non-potential field
(Figure 10; top right) with the potential field (Figure 10; top left) one
can see that the non-zero $\alpha$-parameter induces a helical twist
in the eastern-most sunspot, which causes the field lines emanating
from the sunspot to be rotated by some amount in the anti-clockwise
direction. The bottom panels in Figure 10 clearly demonstrate that the
automated loop tracing code picks out significantly different
geometries for the two cases, so that we expect a significantly
different magnetic field solution, once 3D nonlinear force-free
modeling is attempted.

Using the vertical-current model for non-potential magnetic fields
(i.e., Priest 1982; 2014; Aschwanden 2013a), the azimuthal angle [$\mu$] 
of a helically twisted magnetic field line scales proportionally
to the number $N_{\rm twist}$ of twists and reciprocally to the length $L$ 
of the magnetic-field line, 
\begin{equation}
	\tan{(\mu)} = {2 \pi R N_{twist} \over L} = \alpha R \ ,
\end{equation}
where $R$ is the flux-tube radius, and $\alpha$ is the nonlinear
force-free parameter. From the examples of loop tracing using our
OCCULT-2 code, as shown in Figure 2, we can estimate that the deviation
of tracing from a true magnetic-field line is less than about one pixel
for a loop length of more than 100 pixels. Thus, from $\tan{(\mu)}
\lapprox 0.01$ and $R \approx$ one pixel, {we find a lower limit of the
force-free parameter $\alpha = \tan({\mu})/R \approx 1 \times
10^{-10}$ cm$^{-1}$.} This is an estimate of the sensitivity of our
automated loop-tracing code to the non-potentiality of the magnetic
field. This is commensurable with a range of $\alpha$-values found in
flare-prone regions, \eg $|\alpha| \lapprox 0.02$ arcsec$^{-1}
\approx 3 \times 10^{-10}$ cm$^{-1}$; see Figure 3 in 
Malanushenko \etal 2014).

\section{	Discussion and Conclusions 			}

We developed an automated stereoscopy code that reconstructs the 3D
geometry of coronal loops in a solar active region based on EUV
images, observed with two or three spacecraft located in feasible
orbits, \ie at about 1 AU from the Sun, either ahead or trailing 
behind Earth. In
principle, such a ``blind stereoscopy'' algorithm can be applied to
already existing spacecraft data, such as from SDO/AIA and 
STEREO/EUVI-A(head) and B(ehind), but these existing spacecraft missions have
neither an optimum geometric configuration nor optimum spatial
resolution of the instruments.  We therefore explored the optimum
conditions for a future mission that may place dual spacecraft
anywhere between a near-Earth position and the Lagrangian L4 and L5
points. From our study we arrive at the following conclusions:

\begin{description}

\item{i) \underbar{Spatial Resolution}: The accuracy of stereoscopic
    triangulation scales directly proportional to the spatial
    resolution of the EUV imager for a given space\-craft-separation
    angle, \ie the uncertainty or error $\Delta h_{\rm res} \propto
    \Delta x$ (Equation 11). It is therefore desirable to have the highest
    possible spatial resolution. A combination of requirements of
    full-Sun coverage, high S/N, telemetry, and available
    space-qualified CCD detectors, plus an existing EUV
    Earth-perspective imager already on orbit suggests that a pixel
    size of $0.6\arcsec$ for a {\it $4k \times 4k$} imager as implemented in
    the current SDO/AIA instrument is suitable for the purpose of
    stereoscopic loop tracing.}

\item{ii) \underbar{Number of Spacecraft:} Minimum stereoscopy can be
    performed with a pair of two spacecraft, for instance a spacecraft 
    with one AIA-like telescope at a stereoscopic vantage point, 
    in combination
    with the existing AIA spacecraft in a near-Earth orbit. However, a
    three-spacecraft configuration, such as the existing AIA and two
    twin spacecraft located ahead of and behind the Earth would
    substantially reduce ambiguities in the stereoscopic
    correspondence problem, as well as provide redundancy in case of
    any one instrument failing.}

\item{iii) \underbar{Stereoscopic Correspondence or Ambiguity Problem:} 
    From the simulations of synthetic images of active
    regions we found that the number of stereoscopically corresponding
    loops (detected in an image pair from a spacecraft A and B)
    increases linearly with the spacecraft-separation angle
    $\alpha_s$. This means that larger separation angles lead to
    increased mapping ambiguity.  Thus we should aim for the smallest
    possible angle where stereoscopy is feasible. In contrast,
    large-angle stereoscopy, although feasible, is not at optimum
    conditions. The minimization of the spacecraft-separation angle
    implies also an optimum telemetry rate, since the signal weakens
    with the squared distance to Earth.}

\item{iv) \underbar{Optimum Spacecraft Separation Angle:} The number of
ambiguous loops and thus the uncertainty of stereoscopic triangulated
altitudes increases roughly linearly with the separation angle, 
\ie $\Delta h_{\rm amb} \propto \sin(|\alpha_s|)$ (Equation 12) for small
angles. On the other hand, the uncertainty of stereoscopic triangulated
altitudes decreases reciprocally with the spacecraft-separation angle
due to the limited spatial resolution of the instrument,
\ie $\Delta h_{\rm res} \propto 1/sin{|\alpha_s|}$ (Equation 11). The best
compromise between these two competing effects is at a spacecraft
separation angle $\alpha_{\rm best}$ where the two uncertainties are
comparable, \ie 
$\alpha_{best} \approx \sqrt{2 \Delta x / h_{\rm max}}$ (Equation 15),
which yields $\alpha_{\rm best}=4.6^\circ-6.3^\circ$, for an altitude
limit of $h_{\rm max}=0.1-0.2$ $R_{\odot}$. The altitude limit defines the
vertical extent of the 3D solution space in stereoscopic triangulations.}

\item{v) \underbar{Magnetic Non-Potentiality:} We performed a qualitative
test of how sensitive our automated stereoscopy is to the magnetic topology
of potential and non-potential fields and found that the uncertainties
of automated loop tracing (in the $xy$-plane) has sub-pixel accuracy, 
while the displacements
of loops between a magnetic potential and non-potential field is much
larger (of order $\lapprox 0.1 R_{\odot}$), and thus our automated
stereoscopy code is sufficiently sensitive to changes in the nonlinear
force-free field geometry, down to a nonlinear force-free parameter
of $\alpha \gapprox 10^{-10}$ cm$^{-1}$.}

\end{description}

Our study complements and augments other recent mission concepts that
are proposed as platforms to support space weather research and
operations. One such mission is an L5 Lagrangian Point capability
(Vourlidas, 2015) which focuses on the propagation of CMEs through the
high corona and into the solar wind out to Earth, focusing primarily
on the observation and eventual modeling of the time and velocity of
arrivals of CMEs at Earth. The goals of such an L5 mission would
clearly benefit from a better specification of the Sun--heliosphere field
interface, for instance by adding magnetograph capabilities to
increase the coverage of the solar surface. At this point, the 
proposed concept recognizes that "the entrained magnetic field of an
Earth-directed CME is beyond the reach of current remote-sensing
capabilities."

The goal of the present article is to open up a pathway that addresses
that problem by developing, at least in principle, the methodology to
obtain the information needed for active-region field modeling based
on stereoscopic measurements. Another proposed mission concept, OSCAR
(Strugarek \etal 2015), for example, also explores that through a
perspective from somewhere around L5. OSCAR's premise is that a
suitable stereoscopy angle lies in the range of $\approx 22^\circ -
125^\circ$, based on an earlier work by Aschwanden \etal (2012b).

The earlier study by Aschwanden \etal (2012b) developed a ``quality"
metric for stereoscopy, which included a plausibility argument for the
ambiguity of loop tracing. Here, we quantify a metric for ambiguity
directly from the loop-correspondence algorithm tested here, based on 
simulated coronal images of an active region from different perspectives. 
A result of the blind-stereoscopy algorithm developed here is that
the best performance in resolving the stereoscopic correspondence
ambiguity is found for considerably smaller spacecraft-separation 
angles of $\approx 5^\circ$. At such small angles, the highest 
accuracy is found for coronal stereoscopy to constrain nonlinear
force-free modeling of coronal magnetic fields. This low separation angle,
additionally offers the advantage of a higher telemetry rate for a given
spacecraft and ground-antenna combination.

With these findings we realize that any space-weather
mission concept has to deal with the trade-offs between small-angle
stereoscopy, large-angle CME coverage, and spacecraft telemetry. An
optimum configuration that meets the scientific and operational needs
to study CMEs both in terms of magnetic content and in terms of their
propagation through the heliosphere ({\it cf.}, Schrijver \etal 2015)
would appear to substantially benefit from having a triplet spacecraft
configuration that includes both the L1 and L5 point, and a third
spacecraft about $5^\circ$ away from Earth, equipped with complementary
capabilities to measure the magnetic field, the plasma parameters,
the arrival velocity, and the arrival time of Earth-bound
(geoeffective) CMEs. 

\bigskip 
\acknowledgements Part of the work for automated loop tracing
was supported by the NASA contracts NNG04EA00C of the SDO/AIA
instrument. We also acknowledge Lockheed Martin (LM) independent 
research funding supporting the stereoscopic aspects.

\section*{Disclosure of Potential Conflicts of Interest}

The authors declare that they have no conflicts of interest.

\bibliography{refs}     % file refs.bib

\section*{References} %%% REFERENCES

\def\ref#1{\par\noindent\hangindent1cm {#1}}
\def\aj    {{\it Astron. J. }\ } % Astronomical Journal
\def\aap   {{\it Astron. Astrophys.}\ } % Astronomy and Astrophysics
\def\aar   {{\it Astron. Astrophys. Rev.}\ } % Astronomy and Astrophysics Reviews
\def\apj   {{\it Astrophys. J.}\ } %The Astrophysical Journal
\def\apjl  {{\it Astrophys. J. Lett.}\ } %The Astrophysical Journal
\def\apjss {{\it Astrophys. J. Suppl.}\ } %The Astrophysical Journal
\def\entropy{{\it Entropy }\ } %Entropy 
\def\lrsp  {{\it Living Rev. Solar Phys.}\ } %Living Reviews in Solar Physics
\def\sp    {{\it Solar Phys.}\ } % Solar Physics
\def\ssr   {{\it Space Science Rev.}\ } % Space Science Reviews

\small
\ref{Aschwanden, M.J., Lee, J.K., Gary, G.A., Smith, M.,
	Inhester, B. 1998, \sp {\bf 248}, 359.}
\ref{Aschwanden, M.J., Newmark, J.S., Delaboudini\`{e}re, J.P., 
	Neupert, W.M., Klimchuk, J.A., Gary, G.A., Portier-Fornazzi, F., 
	Zucker,A. 1999, \apj {\bf 515}, 842.}
\ref{Aschwanden, M.J., Alexander, D., Hurlburt, N., Newmark, J.S., 
	Neupert, W.M., Klimchuk, J.A., G.A. Gary 2000, \apj {\bf 531}, 1129.}
\ref{Aschwanden, M.J., W{\"u}lser, J.P., Nitta, N., Lemen,J.
 	2008, \apj {\bf 679}, 827.}
\ref{Aschwanden, M.J. 2010, \sp {\bf 262}, 399.} 
\ref{Aschwanden, M.J., Sandman, A.W. 2010, \aj {\bf 140}, 723.} 
\ref{Aschwanden, M.J. 2011, \lrsp {\bf 8}, 5, DOI: 10.12942/lrsp-2011-5.}
\ref{Aschwanden, M.J., Wuelser, J.P., Nitta, N.V., Lemen, J.R., 
	Schrijver, C.J., DeRosa, M., Malanushenko, A. 
	2012a, \apj {\bf 756}, 124.}
\ref{Aschwanden, M.J., Wuelser, J.P., Nitta, N.V., Lemen, J.R., 
	2012b, \sp {\bf 281}, 101.}
\ref{Aschwanden, M.J. 2013a, \sp {\bf 287}, 323.} 
\ref{Aschwanden, M.J. 2013b, \apj {\bf 763}, 115.} 
\ref{Aschwanden, M.J., De Pontieu, B., Katrukha, E.A.
	2013, \entropy {\bf 15}(8), 3007.}
\ref{Aschwanden, M.J,, Xu, Y., Jing, J. 2014, \apj {\bf 797}, 50.}
\ref{Berton, R. Sakurai, T. 1985, \sp {\bf 96}, 93.}
\ref{DeRosa, M.L., Schrijver, C.J., Barnes, G., Leka, K.D., 
	Lites, B.W., Aschwanden, M.J., Amari,T., Canou, A., \etal
 	2009, \apj {\bf 696}, 1780.}
\ref{Feng, L., Inhester, B., Solanki, S., Wiegelmann, T., 
	Podlipnik, B., Howard, R.A., Wuelser, J.P. 2007, \apj {\bf 671}, L205.}
\ref{Gary, A. 1997, Solar Phys. {\bf 174}, 241.}
\ref{Handy, B.N., Acton, L.W., Kankelborg,C.C., Wolfson, C.J., Akin, D.J., 
	Bruner, M.E., \etal
 	1999, \sp. {\bf 187}, 229.}
\ref{Inhester, B. 2006, {\it Stereoscopic basics for the
	STEREO mission}, ArXiv e-print: astro-ph/0612649.}
\ref{Lemen, J.R., Title, A.M., Akin, D.J., Boerner, P.F., 
	Chou, C., Drake, J.F., Duncan, D.W., Edwards, C.G., \etal
 	2012, \sp {\bf 275}, 17.}
\ref{Malanushenko, A., Schrijver, C.J., DeRosa, M.L., Wheatland, M.S., 
	Gilchrist, S.A. 2012, \apj 756, 153.}
\ref{Malanushenko, A., Schrijver, C.J., DeRosa, M.L., 
	Wheatland, M.S. 2014, \apj {\bf 783}, 102.}
\ref{Pesnell, W.D., Thompson, B.J., Chamberlin, P.C.
 	2011, \sp {\bf 275}, 3.}
\ref{Pevtsov, A.A., Canfield, R.C., Metcalf, T.R.
        1994, \apj {\bf 425}, L117.}  
\ref{Priest, E.R. 1982, {\it Solar Magnetohydrodynamics},
        Reidel, Dordrecht.}
\ref{Priest, E.R. 2014, {\it Magnetohydrodynamics of the Sun},
        Cambridge University Press, Cambridge.}
\ref{Scherrer, P.H., Schou, J., Bush, R. I., Kosovichev, A.G., 
	Bogart, R.S., Hoeksema, J.T., Liu, Y., Duvall, T.L., \etal
 	2012, \sp {\bf 275}, 207.}
\ref{Schrijver, C.J., DeRosa, M.L., Title, A.M., Metcalf, T.R.
        2005, \apj {\bf 628}, 501.}
\ref{Schrijver, C.J., Kauristie, K., Aylward, A.D., Denardini, C.M.,
	Gibson, S.E., Glover, A., Gopalswamy, N., Grandi, M. \etal
	2015, {\it Understanding space weather to shield society: A global
	road map for 2015-2015 commissioned by COSPAR and ILWS},
	Adv. Space Res. {\bf 55}(12), 2745.}
\ref{Strugarek, A., Janitzek, N., Lee, A., L\"oschl, P., Seifert, B.,
	Hoilijoki, S., Kraaikamp, E., Mrigakshi, A.I., \etal
	2015, {\it J. Space Weather Clim.} {\bf 5}, A4.}
\ref{Thompson, W.T., Wei, K. 1010, \sp {\bf 261}, 215.}
\ref{Vourlidas, A. 2015, Space Weather, online-first, DOI 10.1002/2015SW001173.}
\ref{Warren, H.P., Winebarger, A.R. 2007, \apj {\bf 666}, 1245.}
\ref{Welsch, B.T., Christe, S., McTiernan, J.M.
        2011, \sp {\bf 274}, 131.}
\ref{Wiegelmann, T., Inhester, B., Sakurai, T.
 	2006, \sp {\bf 233}, 215.}
\ref{Wiegelmann, T., Sakurai T. 2012, \lrsp {\bf 9}, 5.}
\ref{Wiegelmann, T., Thalmann, J.K., Solanki, K.S. 2014,
        \aar {\bf 22}, 78.}

\end{article}
\end{document}